\RequirePackage{lineno}
\documentclass[showpacs,prl,preprintnumbers,floatfix]{revtex4-1} %one column
\usepackage{graphicx}
\usepackage[english]{babel}
\usepackage{amssymb}
\usepackage{amsbsy}
\usepackage{amsmath}
\usepackage{color}

\begin{document}
\preprint{APS}

\title{Persistent incomplete mixing in reactive flows}

\author{Alexandre M. Tartakovsky$^1$}
\email[E-mail: ]{alexandre.tartakovsky@pnnl.gov}
\author{David Barajas-Solano$^1$}
\affiliation{$^1$Pacific Northwest National Laboratory, Richland, WA 99352, USA}
% \linenumbers

\begin{abstract}
\noindent We present an effective stochastic advection-diffusion-reaction (SADR) model that explains incomplete mixing typically observed in transport with bimolecular reactions. Unlike traditional advection-dispersion-reaction models, the SADR model describes mechanical and diffusive mixing as two separate processes.  In the SADR model, mechanical mixing is driven by random advective velocity with the variance given by the coefficient of mechanical dispersion. The diffusive mixing is modeled as a Fickian diffusion with the effective diffusion coefficient. We demonstrate that the sum of the two coefficients is equal to the dispersion coefficients, but only the  effective diffusion coefficient contributes to the mixing-controlled reactions, indicating that such systems do not get fully mixed at the Representative Elementary Volume scale where the deterministic equations and dispersion coefficient are defined. We use the experimental results of Gramling et al. \cite{Gramling} to show that for transport and bimolecular reactions in porous media, the SADR model is significantly more accurate than the traditional dispersion model, which overestimates the concentration of the reaction product by as much as 60\%.   
\end{abstract}

\maketitle

%%%%%%%%%%%%%%%%%%%%%%
%\section{Introduction}
%%%%%%%%%%%%%%%%%%%%%%

Reactive transport is ubiquitous in natural and man-made systems. At small scales,  reactive transport is dominated by thermally induced velocity fluctuations, and Lagevin-type stochastic diffusion-reaction equation models are well established and have been shown to provide an accurate description of mixing-controlled reactions \cite{risken1984fokker,ReyCardy,PhysRevE.57.2681,ATZBERGER20103474}. In large-scale applications (e.g., atmospheric transport and transport in porous media),  velocity variations are caused by unresolved flow processes (e.g., turbulance; pore-scale flow), and reactive transport is described by deterministic (in the absence of parametric uncertainty) advection-dispersion-reaction  (ADR) equations that tend to overestimate the extent of mixing-controlled reactions \cite{Gramling,Tart2008,TartPRE2010,bolster_pre}. Solutes mix by two distinct mechanisms: 1) molecular diffusion; and 2) mechanical mixing caused by variations in the fluid velocity advecting the solutes. The dispersion theory assumes that solutes are fully mixed within the Representative Elementary Volume and treats both mechanisms as a Fickian diffusive process. We hypothesize that this is the main reason why ADR equations fail for mixing-controlled reactions. There is a complex interplay between these two mixing mechanisms that we investigate by means of a novel stochastic advection-diffusion-reaction, or SADR, model. We demonstrate that the mass of reactive product increases with the average flow velocity and is nearly independent from the magnitude of the velocity fluctuations. We also show that the reactive product mass  grows slowly more than the prediction of the ADR equations. 

\begin{figure}
\centerline{ \includegraphics[width=2.0in]{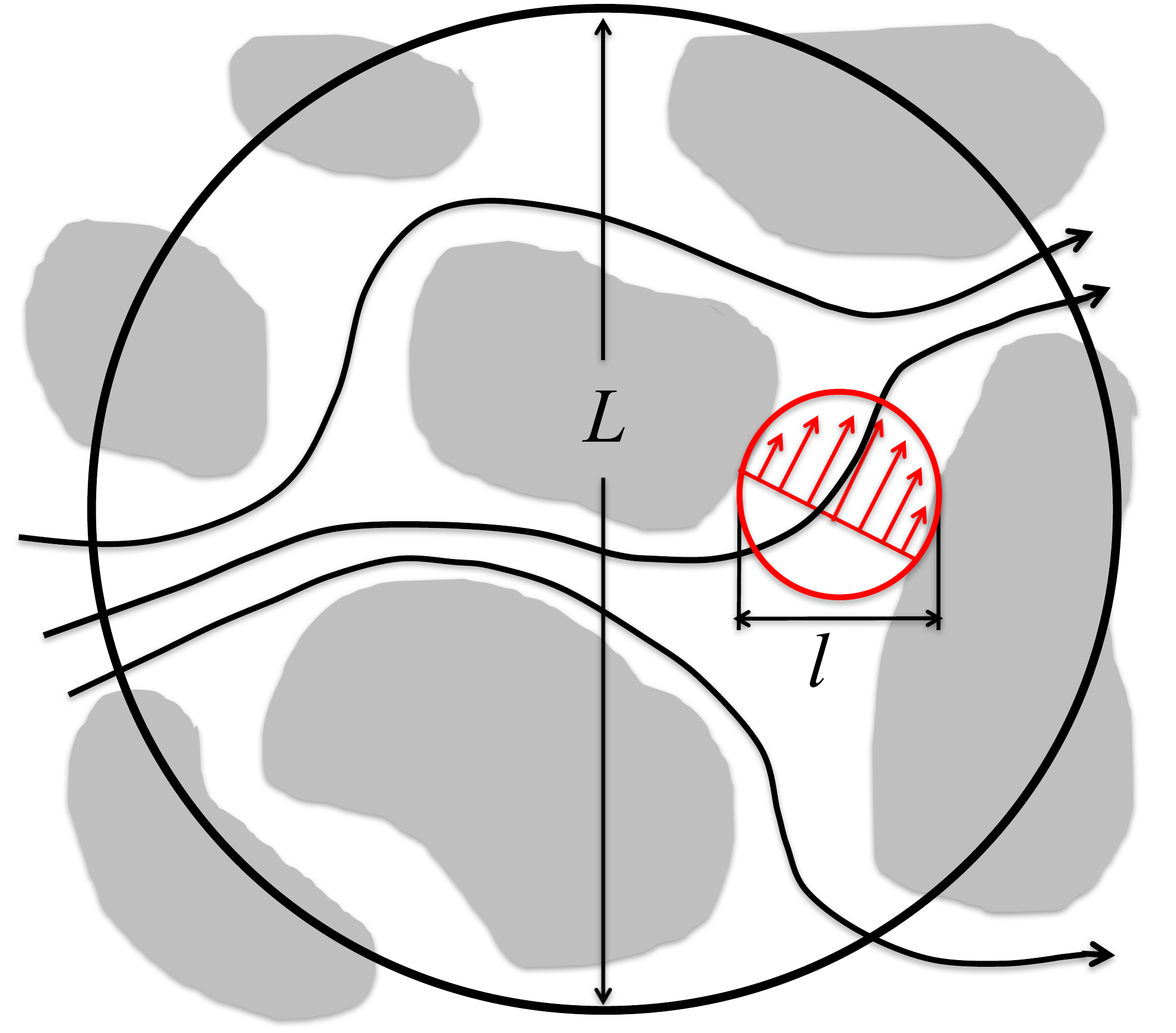} }
\caption{
The SADR model treats  mixing on the scale $l$ as a Fickian process and on the scale $L$ (the Representative Elementary Volume scale) as random advection.}
\label{porous-medium} 
\end{figure}

At the hydrodynamics scale,  an equation for general advection(convection)-diffusion-reaction process  with $N$ species arises naturally from  mass conservation considerations \cite{Pandis,Bearbook}:
\begin{align}\label{diffusion-eq}
\partial_t c_i = d_i  \nabla^2 c_i - \nabla   \cdot 
\left( \mathbf{v}  c_i \right)   - R_i(c_1, ..., c_N) ,
\end{align}
where the advection velocity $\mathbf{v}$ satisfies the Navier-Stokes equations, $c_i$ is the concentration of the $i$th component, $d_i$, is the molecular  diffusion constant, and $R_i$ is the rate of chemical reaction generating (or consuming) the $i$th species. In the case of reactive transport in the atmosphere, ocean, or a river, the advective velocity field  $\mathbf{v}$ is often turbulent. In the case of transport in porous media, $\mathbf{v}$ is the pore-scale velocity defined on the hydrodynamic scale that is much smaller than $l$, the average pore size (Figure \ref{porous-medium}). Under most conditions, direct calculations of  $\mathbf{v}$ are not practical and eq. (\ref{diffusion-eq})  is commonly replaced with the ADR model:
\begin{equation}\label{dispersion-eq}
\partial_t \overline{c}_i =   \mathcal{\mathbf{D}}\nabla^2 \overline{c}_i - \nabla \cdot  \left(  \overline{\mathbf{v}} \,\,  \overline{c}_i \right)   - R_i(\overline{c}_1, ..., \overline{c}_N).
\end{equation}
Here, ${\overline{\mathbf{v}}}$ and $\overline{c_i}$ are the volume or statistical averages of $\mathbf{v}$ and $c_i$, and $\mathcal{\mathbf{D}}$ is the dispersion tensor.  In the case of transport in porous media, the averaging volume has  size $L \gg l$ (Figure \ref{porous-medium}). In the advection-diffusion-reaction equation, the $ \nabla \cdot (\mathbf{v} c_i)$ term models advection and mechanical mixing, and the $d_i  \nabla^2 c_i$ term models diffusive mixing. On the other hand,  in the advection-dispersion-reaction-equation,  both the mechanical and diffusive mixings are modeled by the $\mathcal{\mathbf{D}}\nabla^2 c_i$ term. This is the main reason why the dispersion model often fails to accurately predict mixing-controlled reactions. In  atmospheric and ocean modeling, subgrid models are commonly used to approximate $\mathbf{v}$ and its effect on mechanical mixing \cite{Frederiksen}. With transport in porous media, the state of the art involves using stochastic Lagrangian particle models \cite{Benson,LeBorgne,bolster_pre,Tart2008,Tart2010PRE,DING201356}. While all of these models have been able to accurately describe experiments, their discrete  nature complicates analytical and numerical analysis (e.g., parameters may require calibration, and results of Lagrangian particle models may be resolution-dependent \cite{Benson}).

For clarity of presentation, in the following we focus on transport in porous media, but the model may be applicable to other natural and engineered systems.  
 Here, we present a novel stochastic advection-diffusion-reaction (SADR) model, an effective model defined on the scale $L$.
 In the dispersion model, eq. (\ref{dispersion-eq}), it is assumed that solutes are well mixed on the scale $L$. This mixing is due to large-scale velocity variations within the averaging volume and the Taylor dispersion. The former stems from variations of pore sizes and complex pore geometry, and the latter is the result of velocity variations within individual pores and molecular diffusion.     Chemical reactions, especially of the bimolecular type, create large concentration gradients on scales much smaller than $L$, which violates this assumption. In the SADR model,  we assume that solutes are fully mixed   on the scale $l$ by means of Taylor dispersion. Furthermore, we assume that the ``full'' mixing is a Fickian process with the effective diffusion coefficient $D_m$. Mechanical mixing in the SADR model is caused by random advection velocity. The mean velocity is determined from the effective deterministic (Darcy) flow equations. 
  We formulate the SADR model as the stochastic partial differential equation:
\begin{align}\label{3D-stochastic-diffusion-eq}
\partial_t C_i = \mathbf{D}_m  \nabla^2 C_i - \nabla   \cdot  ( \mathbf{V} C_i)-R_i(C_1, ..., C_N).
\end{align}
Here, $\mathbf{V} = \langle \mathbf{V} \rangle  +  \mathbf{V}' $ is the random advective velocity with mean $\langle \mathbf{V} \rangle$ and random fluctuations $\mathbf{V}'  $. $\langle \mathbf{V} \rangle$  is the average pore velocity that can be found from the Darcy law ($ \langle \mathbf{V} \rangle = -\frac{K}{\theta} \nabla h$, where $K$ is the hydraulic conductivity,  $h$ is the hydraulic head, and $\theta$ is porosity) and continuity equation ($\nabla \cdot (K \nabla h )= 0$), subject to the appropriate initial and boundary conditions.  Random fluctuations $\mathbf{V}'$ have Gaussian probability distribution function, $p_V(\mathbf{u})$, and the covariance function
$
\overline{\mathbf{V}'(\mathbf{x},t) \mathbf{V}'(\mathbf{y},\tau)}
 = 2 \mathbf{D}_d  \rho(\mathbf{x}-\mathbf{y}) \delta(t-\tau),
 $
 where $\delta$ is the Dirac delta function, $\rho$ is a correlation function (e.g., an exponential or Gaussian), and $\mathbf{D}_d$ and $\mathbf{D}_m$ are the mechanical and effective diffusion tensors ( $\mathbf{D}_d + \mathbf{D}_m = \mathbf{D} $) that are defined later.
 The concentration of $i$th species in the porous media is given by the average of $C_i$,  
\begin{equation}
\langle {C}_i(x,t) \rangle  = \int \limits_{-\infty}^{+\infty} C_i(x,t;\mathbf{u}) p_V (\mathbf{u}) d \mathbf{u}.
\end{equation}   
 It can be shown that for non-reactive transport, $\langle {C}_i(x,t) \rangle \equiv \overline{c_i} (x,t)$ \cite{BarajasJUQ}. We will show in this letter that for the non-linear (mixing-controlled) reaction $A+B \rightarrow Y$, the dispersion model overestimates the concentration of reaction product $Y$, i.e., $\langle {Y}(x,t) \rangle < \overline{y} (x,t)$.
 In the following, we use data from a column reactive transport experiment to validate our model. On the  scale $L$, this experiment can be described by the one-dimensional version of eq. (\ref{dispersion-eq}) or (\ref{3D-stochastic-diffusion-eq}).  Because of this, the rest of this paper focuses on the one-dimensional SADR equation:
 \begin{align}\label{stochastic-diffusion-eq}
\partial_t C_i =  D_m  \partial^2_x C_i - V(t) \partial_x  C_i - R_i(C_1, ..., C_N).
\end{align}
The covariance function for one-dimensional divergence-free $V$ is,
\begin{equation}\label{covar-V}
\langle V'(t) V'(\tau) \rangle
 = 2 D_d   \delta(t-\tau).
\end{equation}
 $D_d$ is the part of the dispersion coefficient $\mathcal{D}$ that is independent of  molecular diffusion, and $D_m = \mathcal{D} - D_d$. 
Here, $\mathcal{D}$ is the longitudinal component of $\mathbf{D}$ that, for example, can be determined from a conservative tracer experiment \cite{Bearbook}.

In the rest of this letter, we compare predictions of the SADR model with the experimental data of \cite{Gramling} and demonstrate that the SADR model is more accurate than the dispersion model for a wide range of Peclet numbers $Pe = \langle V \rangle l/ d_m$  (where $d_m$ is the molecular diffusion constant). 
  
\begin{figure}
\centerline{ \includegraphics[width=2.8in]{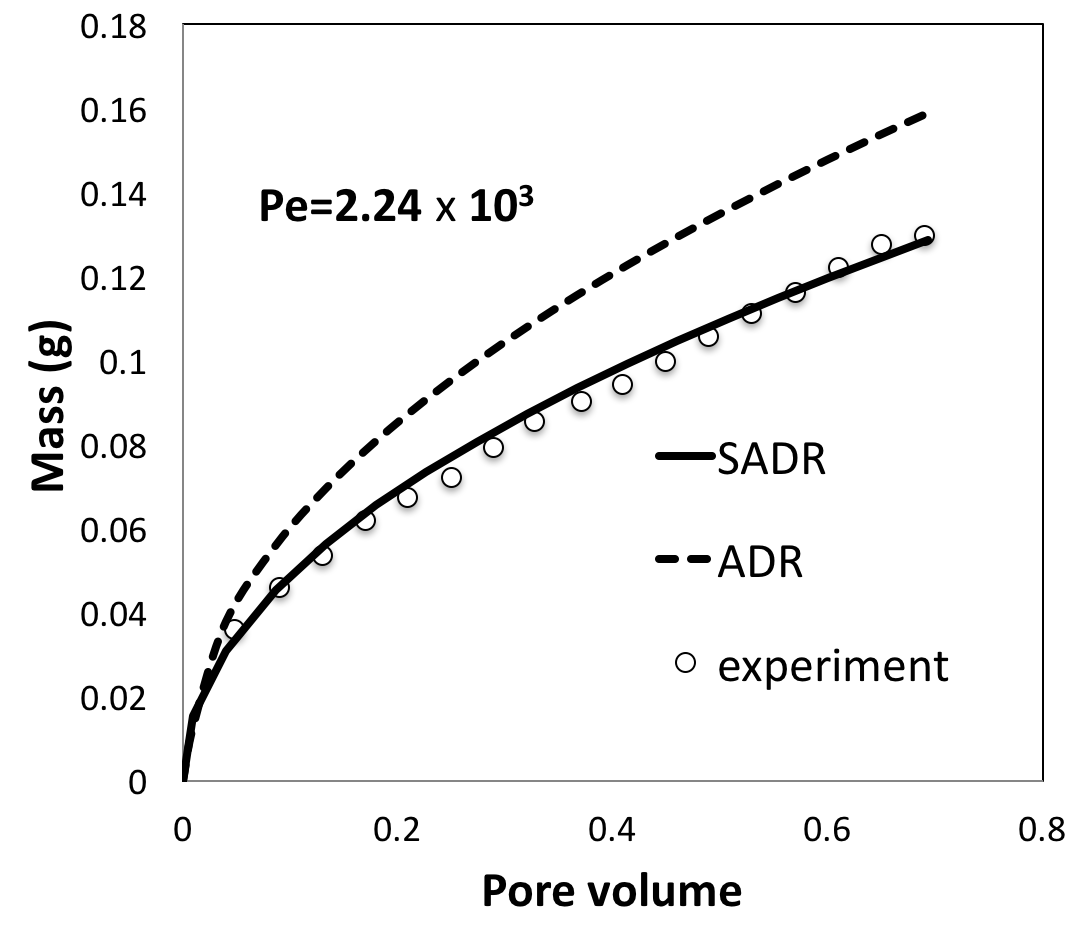} }
\centerline{ \includegraphics[width=2.8in]{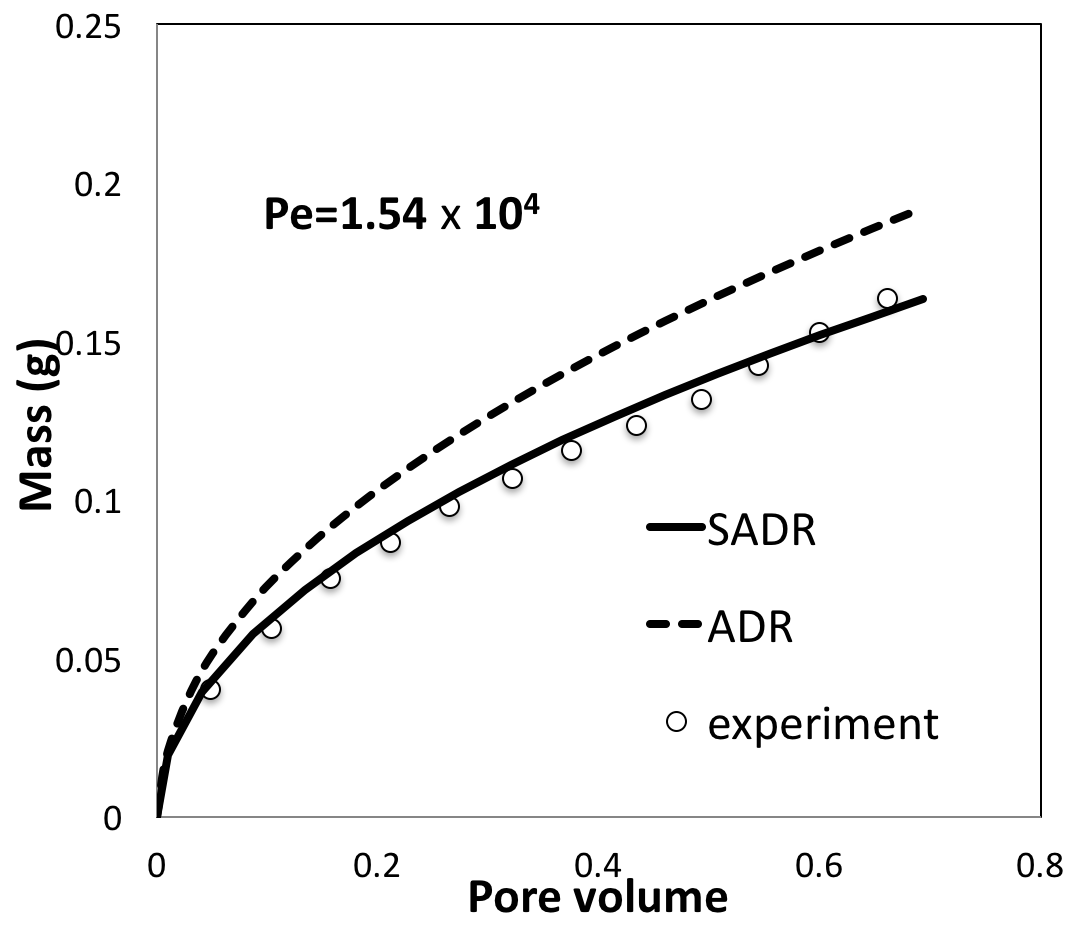} }
\centerline{ \includegraphics[width=2.8in]{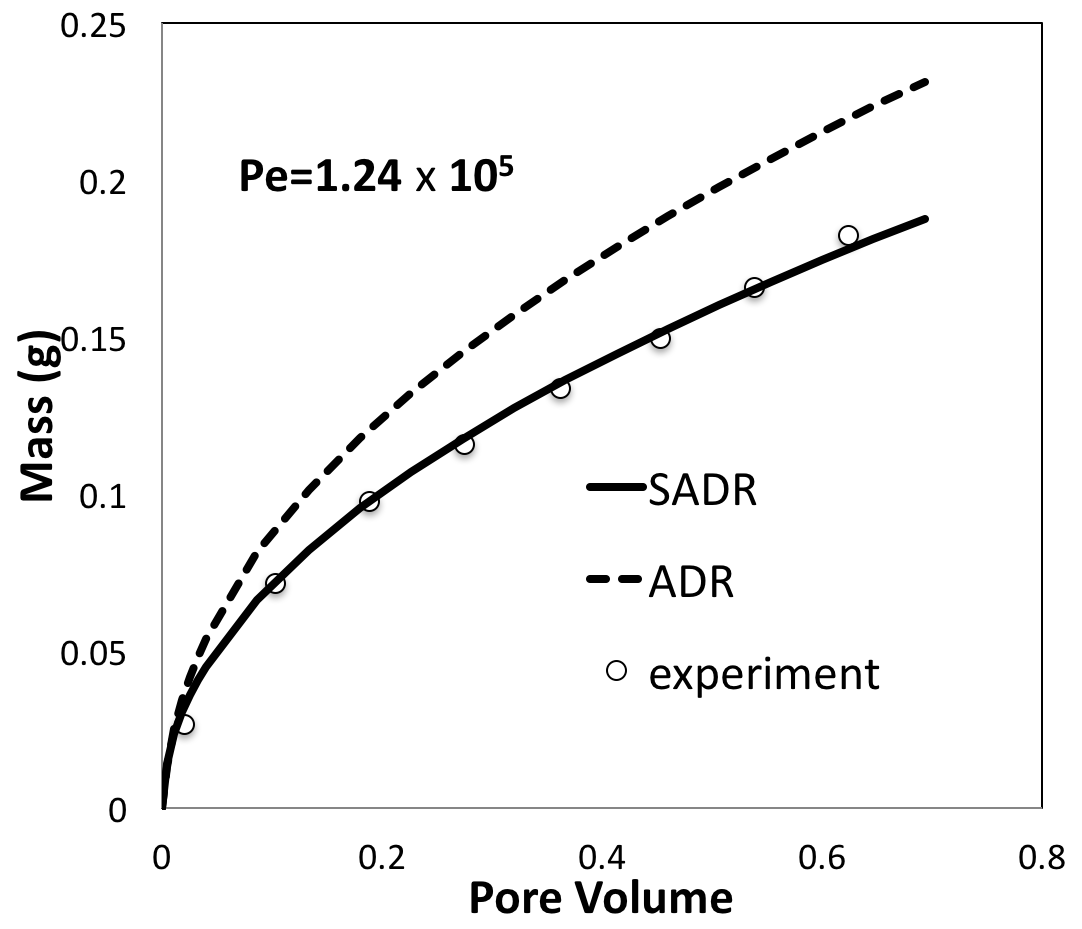} }
\caption{
Total mass of CuEDTA$^{2-},$ predicted by the SADR and advection-dispersion-reaction (ADR) models and measured in the experiments, versus $PV(t) = \langle V \rangle t /\mathcal{L}$ for different $Pe$. 
}
\label{mass}
\end{figure}

\begin{figure}
\centerline{ \includegraphics[width=2.8in]{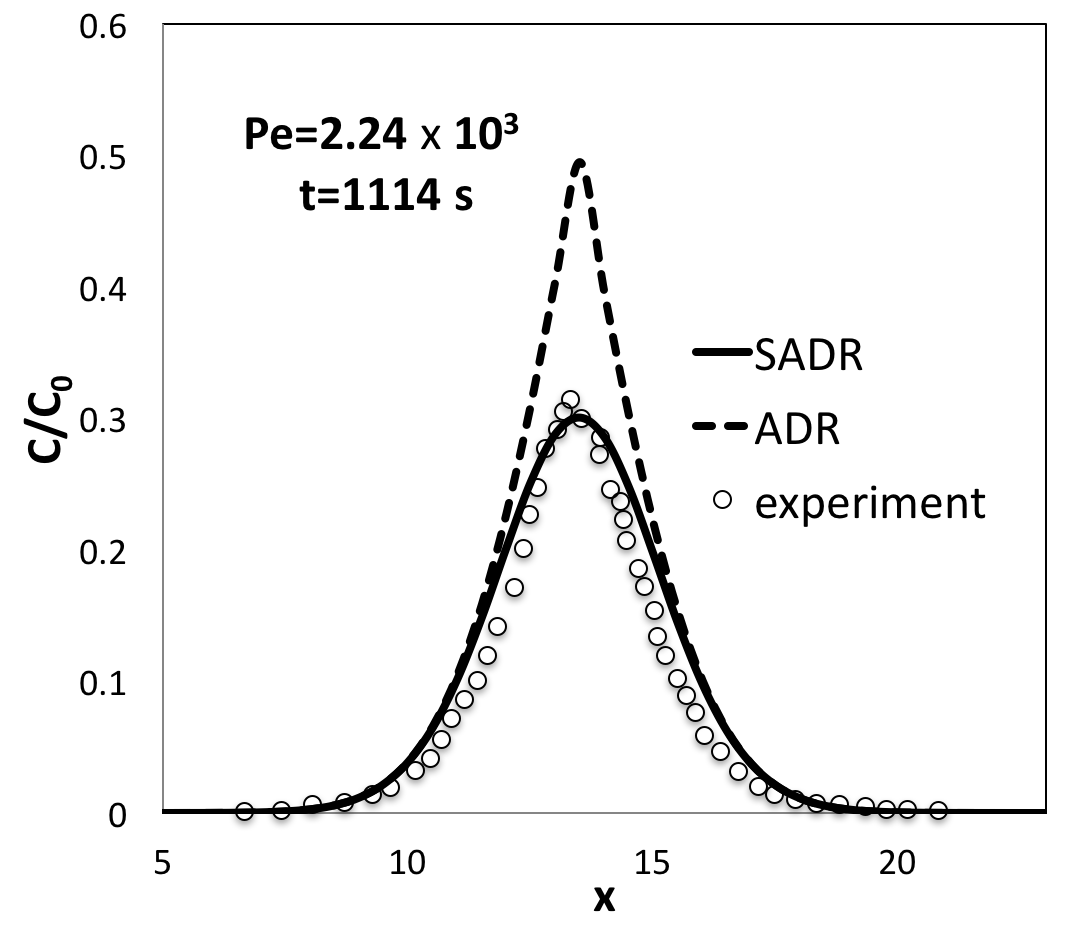} }
\centerline{ \includegraphics[width=2.8in]{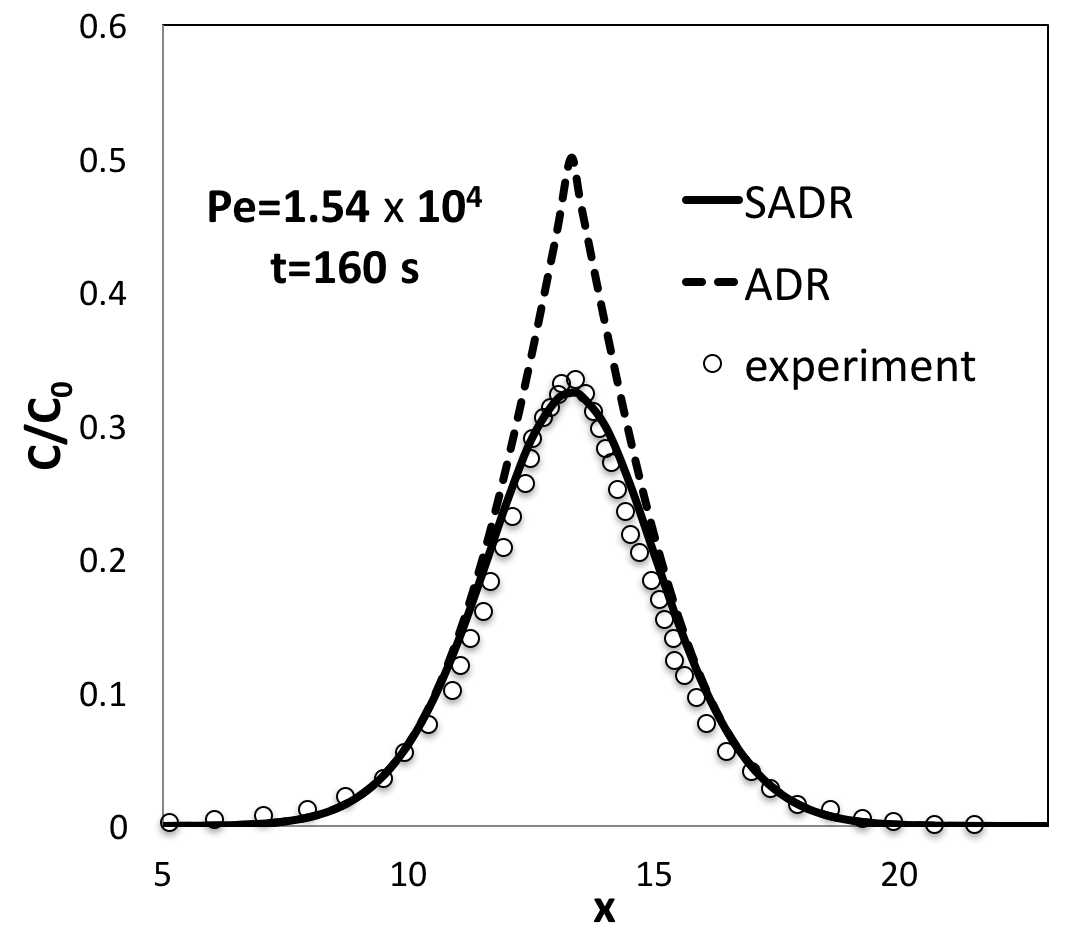} }
\centerline{ \includegraphics[width=2.8in]{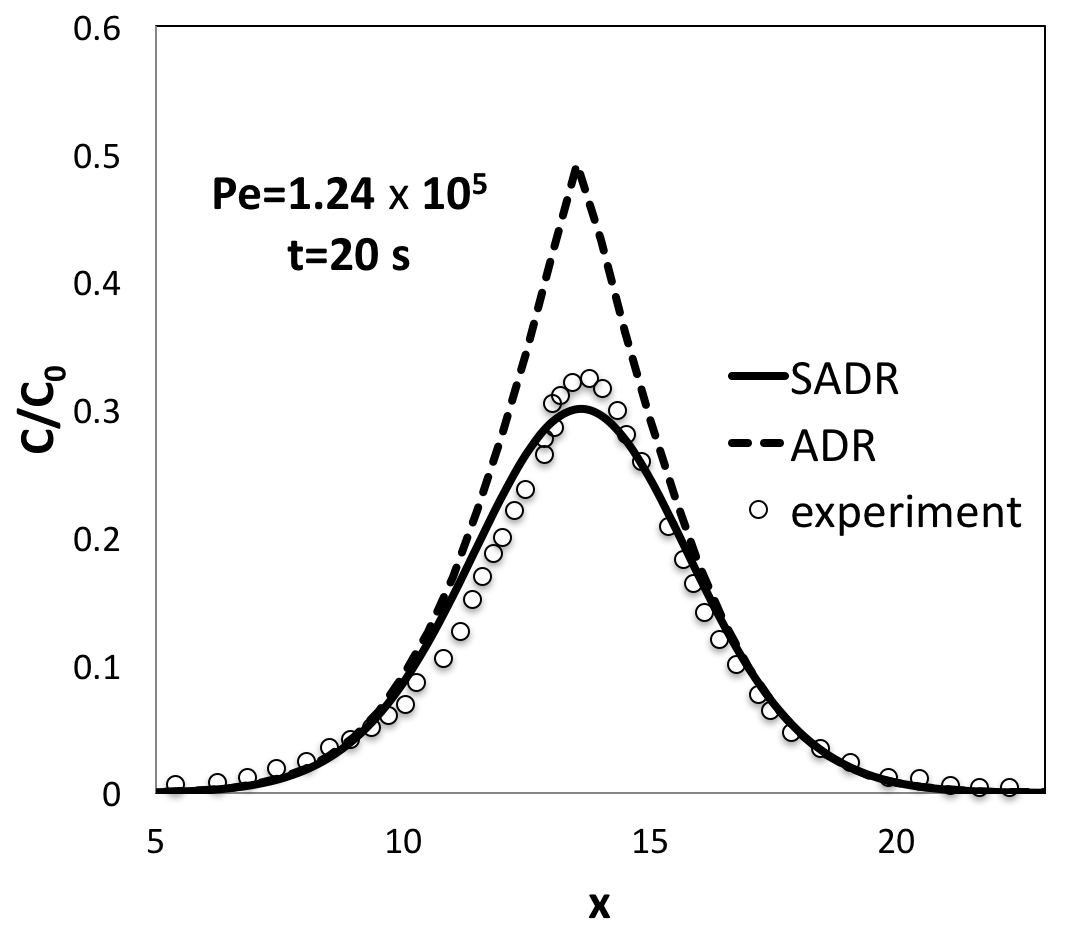} }
\caption{
Concentration of $\text{CuEDTA}^{2-}/C_0,$ predicted by the SADR and ADR models and measured in the experiments, versus $x$ for three different $Pe$. Concentration is shown at the times $t^*$ corresponding to the pore volume  $PV(t) = \langle V \rangle t^*/\mathcal{L} = 0.45$. $\mathcal{L}=30$ cm is the length of the experimental chamber.
}
\label{concentration} 
\end{figure}

A 30-cm long chamber filled with cryolite sand has been used in the experiments. The chamber was initially saturated with a solution containing EDTA$^{4-}$ with  a concentration of $C_0=0.02$ M.  In three different experiments, the  EDTA$^{4-}$ solution was displaced with a solution containing 0.02 M of CuSO$_4$ with flow rates corresponding to $Pe=2.24\times10^3$,  $1.54\times10^4$, and  $=1.24\times10^5$. The instantaneous homogeneous irreversible reaction between the two solutes  formed the reaction product CuEDTA$^{2-}$, whose concentration was measured using a charge-coupled device camera \cite{Gramling}.  Figure \ref{concentration} shows  distributions of the  CuEDTA$^{2-}$ concentration along the chamber, and Figure \ref{mass} depicts the change of the  total mass of  CuEDTA$^{2-}$ in the chamber with time obtained from the experiments. 
According to the dispersion model, the evolution of the concentrations of species in these experiments is described by the one-dimensional form of eq. (\ref{dispersion-eq}) with $R_{\text{CuSO}_4} 
= R_{\text{EDTA}^{4-}} =-R_{\text{CuEDTA}^{2-}} = - k [\text{CuSO}_4] [\text{EDTA}^{4-} ] $, where $k$ is the reaction rate constant. The domain is assumed to be semi-infinite with the prescribed concentration boundary conditions (BCs) at $x=0$
\begin{align}
[\text{CuSO}_4] = C_0,\,[\text{EDTA}^{4-}] = [\text{CuEDTA}^{2-}] = 0
\end{align} 
and free outflow BCs at $x=\infty$
\begin{align}
\partial_x [\text{CuSO}_4] = \partial_x [\text{EDTA}^{4-}]
 = \partial_x [\text{CuEDTA}^{2-}]  =  0.
\end{align} 
The analytical solution for this system of equations is obtained in \cite{Gramling} where the concentration of the reaction product is given by 
\begin{equation}\label{dispersion-cons}
[\text{CuEDTA}^{2-}] (x,t) = \frac{C_0}2 \text{erfc} \left[
\frac{|x - \overline{v}t |  } {2\sqrt{\mathcal{D} t}}  
 \right],
\end{equation}
and the total mass of the reactive product is equal to
\begin{equation}\label{dispersion-mass-exact}
\frac{M(t)}{C_0} =\frac{1}{C_0} \int \limits_0^{+\infty} [\text{CuEDTA}^{2-}] (x,t)dx 
 =2 \sqrt{\frac{\mathcal{D} t}{\pi}}
 - \frac{\sqrt{\mathcal{D} t}}{\sqrt{\pi}} \exp \left[-\frac{ \overline{v}^2 t}{4 \mathcal{D} }\right]
 +
 \frac12 \overline{v}t  \text{erfc} \left[ \frac{ \overline{v} t}{2\sqrt{\mathcal{D} t}} \right] .
\end{equation}

For the ratio of advection length to the diffusion length greater than one,  $ \frac{\overline{v} t}{\sqrt{\mathcal{D} t}} > 1$, the expression for the total mass can be approximated as 
\begin{equation}\label{dispersion-mass}
\frac{M(t)}{C_0} =\frac{1}{C_0} \int \limits_0^{+\infty} [\text{CuEDTA}^{2-}] (x,t)dx 
 =2 \sqrt{\frac{\mathcal{D} t}{\pi}}.
\end{equation}

Figures  \ref{mass} and \ref{concentration} compare these analytical solutions with the experimental data and show that the dispersion model overpredicts the concentration and mass of the reaction product by more than 20\% for all considered $Pe$.  The values of $\overline{v}$ and $\mathcal{D}$ in eqs. (\ref{dispersion-cons}) and (\ref{dispersion-mass})  were measured from  conservative tracer experiments in \cite{Gramling} and are given in Table \ref{parameters}.

The SADR model for these experiments takes the form:
$
\partial_t A = D_m \partial^2_x A - V(t) \partial_x A - kAB,
$
$
\partial_t B = D_m \partial^2_x B - V(t) \partial_x B-kAB,
$
and
$
\partial_t C = D_m \partial^2_x C - V(t) \partial_t C + kAB.
$
The concentrations of the species are given by $[\text{CuSO}_4] = \langle A \rangle$,  
$[\text{EDTA}^{4-}] =  \langle B \rangle $, and $[\text{CuEDTA}^{2-}] = \langle C \rangle $.
Using the change of variables, $y = x- \int_0^tV(t')dt' =x- \langle V \rangle t - \int_0^tV'(t')dt' = x -  \langle V \rangle t - \Gamma(t)$, we rewrite these equations as
$
\partial_t A = D_m \partial^2_y A  -kAB,
$
$
\partial_t B =D_m \partial^2_y B - kAB
$
and
$
\partial_t C = D_m \partial^2_y C + kAB.
$
Under the assumption of instantaneous reaction, these equations have an approximate analytical solution \cite{Gramling}, which for the variable $C$ in terms of the old coordinate $x$ takes the form:
\begin{equation}
C(x,t;\Gamma) = \frac{C_0}2 \text{erfc} \left[
\frac{|x - \langle V \rangle t - \Gamma (t) |  } {2\sqrt{D_m t} }  
 \right].
\end{equation}
Here, $\Gamma = \int_0^t V'(t')dt'$ is the random variable. Given that $V'$ has Gaussian probability density with zero-mean and the covariance given by eq. (\ref{covar-V}), $\Gamma$ also has the Gaussian probability density, $p_\Gamma$, with  the variance $2D_d t$,      
\begin{equation}
p_\Gamma (\gamma) = \frac1{\sqrt{4\pi D_{d} t}}\exp \left( -\frac{\gamma^2}{4 D_{d} t} \right).
\end{equation}
Finally, the reactive product concentration $[\text{CuEDTA}^{2-}]$ is given by 
\begin{equation}\label{average-concentration}
\langle {C}(x,t) \rangle   = \int \limits_{-\infty}^{+\infty} C(x,t;\gamma) p_\Gamma (\gamma) d \gamma,
\end{equation}
and the mass of $[\text{CuEDTA}^{2-}]$ is found as
\begin{eqnarray}\label{mass-diffusive-exact}
\frac{M(t)}{C_0} &=& \frac1{C_0} \int \limits_0^{+\infty} \langle {C}(x,t) \rangle dx 
=  2 \sqrt{ \frac{D_m t }{\pi} }  -  
\sqrt{\frac{\mathcal{D}t}{\pi}}  \exp \left[ -\frac{\langle V \rangle^2 t }{ 4\mathcal{D}} \right] 
+ \frac12 \langle V \rangle t  \text{erfc} \left[ \frac{\langle V \rangle t }{2\sqrt{\mathcal{D}t}}  \right] .
\end{eqnarray}

For large time, i.e., $t > \mathcal{D}/\langle V \rangle^2$, this expression reduces to 
 \begin{eqnarray}\label{mass-diffusive}
\frac{M(t)}{C_0} =  2 \sqrt{ \frac{D_m t }{\pi} }. 
\end{eqnarray}

Equations (\ref{mass-diffusive-exact}) and (\ref{mass-diffusive}) are the main theoretical results of this work. The comparison of  eqs. (\ref{mass-diffusive-exact}) and (\ref{mass-diffusive}) with eqs. (\ref{dispersion-mass-exact}) and (\ref{dispersion-mass}) reveals that: 1) the mechanical mixing contributes to the  reaction only at an early time, i.e., $t  < \mathcal{D}/\langle V \rangle^2$; 2) at late time,  only $D_m< \mathcal{D}$ contributes to the reaction; and 3) the system never gets fully mixed as suggested by the advection-dispersion model.     
 
We find  $D_m$ by fitting eq. (\ref{mass-diffusive}) to the reactive product mass obtained in the experiments (see Figure \ref{mass}) and report these values in Table \ref{parameters}. Next, we validate the model by computing the average concentration from eq. (\ref{average-concentration}) and comparing it with the concentration found from the experiment. Figure \ref{concentration} shows that the concentration found from the SADR equations agrees with the experimental values for all considered $Pe$, while the  dispersion model (eq. (\ref{dispersion-cons})) overestimates the maximum concentration by as much as 60\%.

%%%%%%%%%%%%%%%%%%%%%%
%
% Empirical model for $D_m$
%
%%%%%%%%%%%%%%%%%%%%%%
Finally, we propose an empirical approach to estimate $D_m$ and $D_d$ for cases when these coefficients cannot be readily obtained from experiments (as described herein). We assume that the dispersion coefficient as a function of time is available from a conservative transport experiment and/or follow the  relationship \cite{Coelho}:
\begin{equation}\label{D-Pe-relationship}
\mathcal{D} = d_m \alpha Pe^{1+\beta},
\end{equation}
 where  $0<\alpha<0.5$ and $0<\beta<0.5$ are experimentally determined constants.

 Expanding eq. (\ref{D-Pe-relationship}) in Taylor series in powers of $\beta$ yields
\begin{equation}\label{d-expansion}
 \mathcal{D} = d_m \alpha Pe \left[
  1 + \beta  \log Pe +  \frac{\beta^2  (\log Pe)^2}{2!} + o(\beta^3)
  \right].
\end{equation}
Only the zero-order (in $\beta$) term in eq. (\ref{d-expansion}) is independent of molecular diffusion and, hence, contributes to the mechanical dispersion. Therefore, we define $D_d$ and $D_m$ as
\begin{equation}\label{d-m}
 D_d  = d_m \alpha Pe 
\end{equation}
and
\begin{equation}\label{d-d}
D_m =  \mathcal{D}-D_d.
\end{equation}
For typical values of $\beta$ ($\approx 0.25$), keeping terms up to 5th order in the Taylor expansion results in the error being less than 1\%. In the following, we will refer to $D_m$ and $D_d$ defined by eqs. (\ref{d-m}) and (\ref{d-d}) as "first-order coefficients".   In three-dimensional problems, eqs. (\ref{D-Pe-relationship}) - (\ref{d-d}) can be used to define both longitudinal and transverse components of the dispersion, mechanical dispersion, and effective diffusion tensors. The coefficients $\alpha$ and $\beta$ are different for longitudinal and  transverse components of the tensors and could be determined from conservative tracer experiments. 

\begin{figure}
\centerline{ \includegraphics[width=2.8in]{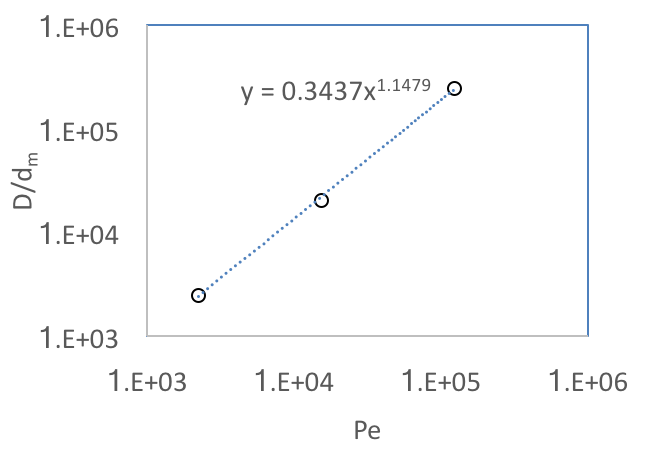} }
\caption{
Fit of eq. (\ref{d-expansion}) to the experimentally determined dispersion coefficient normalized with the molecular diffusion coefficient $\mathcal{D}/d_m$.
}
\label{dispesion_coef} 
\end{figure}

For the considered experiment, $\alpha=0.34$ and $\beta = 0.14$ are found by fitting eq. (\ref{D-Pe-relationship}) to the experimentally determined values of $\mathcal{D}$ and $Pe$ (see Figure \ref{dispesion_coef}). The resulting first-order $D_m$ and $D_d$ are given in Table \ref{parameters}, where we assumed that all reactive species have the same molecular diffusion coefficient, which was set to $d_m = 7.02\times10^{-7}$ (the diffusion coefficient for CuEDTA$^{2-}$).

\begin{figure}
\centerline{ \includegraphics[width=2.8in]{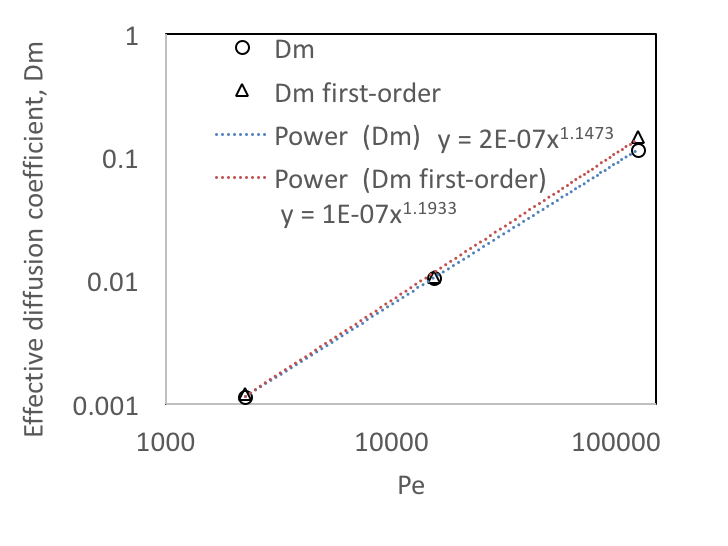} }
\caption{
The ``first-order'' effective diffusion coefficient given by eq. (\ref{d-m}) and obtained by fitting eq. (\ref{mass-diffusive}) to the experimental data versus $Pe$.
}
\label{effective-diff_coef} 
\end{figure}

Figure \ref{effective-diff_coef} shows the ``first-order'' effective diffusion coefficient and the effective diffusion coefficient $D_m$ obtained by fitting eq. (\ref{mass-diffusive}) to the experimental data. The values of these coefficients are given in Table \ref{parameters}. Notably, eq. ({\ref{d-m}}) gives a good estimate of $D_m$, especially for smaller $Pe$. It is interesting that both $\mathcal{D}$ and $D_m$ scale as $Pe^\beta$.

\begin{table}
\caption{Parameters and errors in the dispersion and SADR models. $l=0.13$ cm and $d_m=7.02\times10^{-7}$ cm$^2$/s.}
\begin{tabular}{|c|c|c|c|}
	\hline
$Pe$                                      & $2.24\times10^3    $  &$1.54\times10^4$      & $1.24\times10^5$     \\
$ \langle V \rangle = \overline{v}$ (cm/s) &$1.21\times10^{-2} $ & $8.32\times10^{-2}$ & $6.70\times10^{-1} $\\
$ \mathcal{D}$ (cm$^2$/s) & $1.75\times10^{-3}$ & $1.45\times10^{-2}$ & $1.75\times10^{-1}$ \\
$D_m$ (cm$^2$/s)   & 0.0012 & 0.011 & 0.115 \\
$D_d$   (cm$^2$/s)   & 0.0006 &0.004  & 0.06\\
1st-order $D_m$ (cm$^2$/s)   & 0.0012 & 0.011 & 0.145 \\
1st-order $D_d$   (cm$^2$/s)   & 0.0006 &0.004  & 0.03\\
	\hline
\end{tabular}
\label{parameters}
\end{table}

%%%%%%%%%%%%%%%%%%%%%%%%%%%%%%%%%%%%%%%%%%%%%%%%%%%%%%%
%%%%%%%%%%%%%%%%%%%%%      CONCLUSION       %%%%%%%%%%%%%%%%%%%%%%%%%%%
%%%%%%%%%%%%%%%%%%%%%%%%%%%%%%%%%%%%%%%%%%%%%%%%%%%%%%%
%\section{Conclusion}
%\label{sec:conclusion}
{\it In summary}, we have proposed a new SADR model for multicomponent reactive transport in porous media.  Using experimental data, we have demonstrated that the SADR model is significantly more accurate than the classical dispersion model.  
Our model treats mixing as a combination of purely mechanical and diffusive mixing characterized by the mechanical and effective diffusing coefficients. The sum of these two coefficients is equal to the dispersion coefficients. The model predicts that only the effective diffusion coefficient contributes to the reaction, which is the reason the dispersion model overestimates the mass of the reactive product. 
The important feature of the SADR model is that it is defined in terms of parameters and variables (such as average pore velocity and dispersion coefficient), which can be measured or estimated on the continuum (Darcy) scale.  No sub-scale (pore-scale) information or simulations are needed to parameterize the SADR model. The concept of separating mechanical mixing and diffusive mixings is not limited to transport in porous media, and the SADR model can be extended to other fields.   

\noindent
{\bf Acknowledgments\\}
This research  was supported by the Advanced Scientific Computing Research Program of the U.S. Department of Energy (DOE) Office of Science. Pacific Northwest National Laboratory is operated by Battelle for the DOE under Contract DE-AC05-76RL01830.

\bibliographystyle{apsrev}
%\bibliography{stochastic-ad-diffusion-reaction} 

\begin{thebibliography}{17}
\expandafter\ifx\csname natexlab\endcsname\relax\def\natexlab#1{#1}\fi
\expandafter\ifx\csname bibnamefont\endcsname\relax
  \def\bibnamefont#1{#1}\fi
\expandafter\ifx\csname bibfnamefont\endcsname\relax
  \def\bibfnamefont#1{#1}\fi
\expandafter\ifx\csname citenamefont\endcsname\relax
  \def\citenamefont#1{#1}\fi
\expandafter\ifx\csname url\endcsname\relax
  \def\url#1{\texttt{#1}}\fi
\expandafter\ifx\csname urlprefix\endcsname\relax\def\urlprefix{URL }\fi
\providecommand{\bibinfo}[2]{#2}
\providecommand{\eprint}[2][]{\url{#2}}

\bibitem[{\citenamefont{Gramling et~al.}(2002)\citenamefont{Gramling, Harvey,
  and Meigs}}]{Gramling}
\bibinfo{author}{\bibfnamefont{C.~M.} \bibnamefont{Gramling}},
  \bibinfo{author}{\bibfnamefont{C.~F.} \bibnamefont{Harvey}},
  \bibnamefont{and} \bibinfo{author}{\bibfnamefont{L.~C.} \bibnamefont{Meigs}},
  \bibinfo{journal}{Environ. Sci. Technol.} \textbf{\bibinfo{volume}{36}},
  \bibinfo{pages}{2508} (\bibinfo{year}{2002}).

\bibitem[{\citenamefont{Risken}(1984)}]{risken1984fokker}
\bibinfo{author}{\bibfnamefont{H.}~\bibnamefont{Risken}},
  \emph{\bibinfo{title}{Fokker-Planck Equation}}
  (\bibinfo{publisher}{Springer}, \bibinfo{year}{1984}).

\bibitem[{\citenamefont{Rey and Cardy}(1999)}]{ReyCardy}
\bibinfo{author}{\bibfnamefont{P.-A.} \bibnamefont{Rey}} \bibnamefont{and}
  \bibinfo{author}{\bibfnamefont{J.}~\bibnamefont{Cardy}},
  \bibinfo{journal}{Journal of Physics A: Mathematical and General}
  \textbf{\bibinfo{volume}{32}}, \bibinfo{pages}{1585} (\bibinfo{year}{1999}).

\bibitem[{\citenamefont{Deem and Park}(1998)}]{PhysRevE.57.2681}
\bibinfo{author}{\bibfnamefont{M.~W.} \bibnamefont{Deem}} \bibnamefont{and}
  \bibinfo{author}{\bibfnamefont{J.-M.} \bibnamefont{Park}},
  \bibinfo{journal}{Phys. Rev. E} \textbf{\bibinfo{volume}{57}},
  \bibinfo{pages}{2681} (\bibinfo{year}{1998}),
  \urlprefix\url{https://link.aps.org/doi/10.1103/PhysRevE.57.2681}.

\bibitem[{\citenamefont{Atzberger}(2010)}]{ATZBERGER20103474}
\bibinfo{author}{\bibfnamefont{P.~J.} \bibnamefont{Atzberger}},
  \bibinfo{journal}{Journal of Computational Physics}
  \textbf{\bibinfo{volume}{229}}, \bibinfo{pages}{3474 }
  (\bibinfo{year}{2010}), ISSN \bibinfo{issn}{0021-9991},
  \urlprefix\url{http://www.sciencedirect.com/science/article/pii/S0021999110000276}.

\bibitem[{\citenamefont{Tartakovsky et~al.}(2008)\citenamefont{Tartakovsky,
  Tartakovsky, and Meakin}}]{Tart2008}
\bibinfo{author}{\bibfnamefont{A.~M.} \bibnamefont{Tartakovsky}},
  \bibinfo{author}{\bibfnamefont{D.~M.} \bibnamefont{Tartakovsky}},
  \bibnamefont{and} \bibinfo{author}{\bibfnamefont{P.}~\bibnamefont{Meakin}},
  \bibinfo{journal}{Physical Review Letters} \textbf{\bibinfo{volume}{101}},
  \bibinfo{pages}{044502} (\bibinfo{year}{2008}).

\bibitem[{\citenamefont{Tartakovsky}(2010{\natexlab{a}})}]{TartPRE2010}
\bibinfo{author}{\bibfnamefont{A.~M.} \bibnamefont{Tartakovsky}},
  \bibinfo{journal}{Physical Review E} \textbf{\bibinfo{volume}{82}}
  (\bibinfo{year}{2010}{\natexlab{a}}).

\bibitem[{\citenamefont{Bolster et~al.}(2010)\citenamefont{Bolster, Benson,
  Borgne, and Dentz}}]{bolster_pre}
\bibinfo{author}{\bibfnamefont{D.}~\bibnamefont{Bolster}},
  \bibinfo{author}{\bibfnamefont{D.}~\bibnamefont{Benson}},
  \bibinfo{author}{\bibfnamefont{T.~L.} \bibnamefont{Borgne}},
  \bibnamefont{and} \bibinfo{author}{\bibfnamefont{M.}~\bibnamefont{Dentz}},
  \bibinfo{journal}{Physical Review E} \textbf{\bibinfo{volume}{82}},
  \bibinfo{pages}{021119} (\bibinfo{year}{2010}).

\bibitem[{\citenamefont{Pandis and Seinfeld}(2006)}]{Pandis}
\bibinfo{author}{\bibfnamefont{S.~N.} \bibnamefont{Pandis}} \bibnamefont{and}
  \bibinfo{author}{\bibfnamefont{J.~H.} \bibnamefont{Seinfeld}},
  \emph{\bibinfo{title}{Atmospheric chemistry and physics: from air pollution
  to climate change}} (\bibinfo{publisher}{J. Wiley}, \bibinfo{year}{2006}).

\bibitem[{\citenamefont{Bear}(1988)}]{Bearbook}
\bibinfo{author}{\bibfnamefont{J.}~\bibnamefont{Bear}},
  \emph{\bibinfo{title}{Dynamics of fluids in porous media}}
  (\bibinfo{publisher}{Dover Publications}, \bibinfo{address}{New York},
  \bibinfo{year}{1988}).

\bibitem[{\citenamefont{Frederiksen}(2012)}]{Frederiksen}
\bibinfo{author}{\bibfnamefont{J.~S.} \bibnamefont{Frederiksen}},
  \bibinfo{journal}{Entropy} \textbf{\bibinfo{volume}{14}}, \bibinfo{pages}{32}
  (\bibinfo{year}{2012}).

\bibitem[{\citenamefont{Benson and Meerschaert}(2009)}]{Benson}
\bibinfo{author}{\bibfnamefont{D.}~\bibnamefont{Benson}} \bibnamefont{and}
  \bibinfo{author}{\bibfnamefont{M.}~\bibnamefont{Meerschaert}},
  \bibinfo{journal}{Water Resources Research} \textbf{\bibinfo{volume}{44}},
  \bibinfo{pages}{W12201} (\bibinfo{year}{2009}).

\bibitem[{\citenamefont{LeBorgne et~al.}(2008)\citenamefont{LeBorgne, Dentz,
  and Carrera}}]{LeBorgne}
\bibinfo{author}{\bibfnamefont{T.}~\bibnamefont{LeBorgne}},
  \bibinfo{author}{\bibfnamefont{M.}~\bibnamefont{Dentz}}, \bibnamefont{and}
  \bibinfo{author}{\bibfnamefont{J.}~\bibnamefont{Carrera}},
  \bibinfo{journal}{Phys Rev Lett} \textbf{\bibinfo{volume}{101}},
  \bibinfo{pages}{090601} (\bibinfo{year}{2008}).

\bibitem[{\citenamefont{Tartakovsky}(2010{\natexlab{b}})}]{Tart2010PRE}
\bibinfo{author}{\bibfnamefont{A.~M.} \bibnamefont{Tartakovsky}},
  \bibinfo{journal}{Physical Review E} \textbf{\bibinfo{volume}{82}},
  \bibinfo{pages}{026302} (\bibinfo{year}{2010}{\natexlab{b}}).

\bibitem[{\citenamefont{Ding et~al.}(2013)\citenamefont{Ding, Benson, Paster,
  and Bolster}}]{DING201356}
\bibinfo{author}{\bibfnamefont{D.}~\bibnamefont{Ding}},
  \bibinfo{author}{\bibfnamefont{D.~A.} \bibnamefont{Benson}},
  \bibinfo{author}{\bibfnamefont{A.}~\bibnamefont{Paster}}, \bibnamefont{and}
  \bibinfo{author}{\bibfnamefont{D.}~\bibnamefont{Bolster}},
  \bibinfo{journal}{Advances in Water Resources} \textbf{\bibinfo{volume}{53}},
  \bibinfo{pages}{56 } (\bibinfo{year}{2013}), ISSN \bibinfo{issn}{0309-1708},
  \urlprefix\url{http://www.sciencedirect.com/science/article/pii/S030917081200276X}.

\bibitem[{\citenamefont{Barajas-Solano and M.Tartakovsky}(2017)}]{BarajasJUQ}
\bibinfo{author}{\bibfnamefont{D.}~\bibnamefont{Barajas-Solano}}
  \bibnamefont{and}
  \bibinfo{author}{\bibfnamefont{A.}~\bibnamefont{M.Tartakovsky}},
  \bibinfo{journal}{Journal of Uncertainty Quantification}
  \textbf{\bibinfo{volume}{M110916}} (\bibinfo{year}{2017}).

\bibitem[{\citenamefont{Coelho et~al.}(1997)\citenamefont{Coelho, Thovert, and
  Adler}}]{Coelho}
\bibinfo{author}{\bibfnamefont{D.}~\bibnamefont{Coelho}},
  \bibinfo{author}{\bibfnamefont{J.-F.} \bibnamefont{Thovert}},
  \bibnamefont{and} \bibinfo{author}{\bibfnamefont{P.~M.} \bibnamefont{Adler}},
  \bibinfo{journal}{Phys. Rev. E} \textbf{\bibinfo{volume}{55}},
  \bibinfo{pages}{1959} (\bibinfo{year}{1997}).

\end{thebibliography}

\end{document}